\documentclass[sigconf, nonacm]{acmart}

\usepackage{mathtools}

\DeclarePairedDelimiter{\set}{\{}{\}}

\usepackage{balance}

\AtBeginDocument{%
  \providecommand\BibTeX{{%
    \normalfont B\kern-0.5em{\scshape i\kern-0.25em b}\kern-0.8em\TeX}}}
\usepackage{color, xcolor}
\usepackage{balance}
\usepackage{subfigure}

\begin{document}

\title[Professional Network Matters: Connections Empower Person-Job Fit]{Professional Network Matters:\\ Connections Empower Person-Job Fit}

\author{Hao Chen}
\email{haochen@mail.bnu.edu.cn}
\authornote{This work was done when they were interns at LinkedIn.}
\affiliation{\institution{Beijing Normal University}
    \city{Beijing}
    \country{China}
}

\author{Lun Du}
\email{lun.du@microsoft.com}
\authornote{Corresponding author}
\affiliation{\institution{Microsoft Research}
    \city{Beijing}
    \country{China}
}

\author{Yuxuan Lu}
\email{lu.yuxuan@northeastern.edu}
\authornotemark[1]
\affiliation{\institution{Northeastern University}
    \city{Boston}
    \country{USA}
}

\author{Qiang Fu}
\email{qifu@microsoft.com}
\affiliation{\institution{Microsoft Research}
    \city{Beijing}
    \country{China}
}

\author{Xu Chen}
\email{xu.chen@microsoft.com}
\affiliation{\institution{Microsoft Research}
    \city{Beijing}
    \country{China}
}

\author{Shi Han}
\email{shihan@microsoft.com}
\affiliation{\institution{Microsoft Research}
    \city{Beijing}
    \country{China}
}

\author{Yanbin Kang}
\email{ybkang@linkedin.com}
\affiliation{\institution{LinkedIn Corporation}
    \city{Beijing}
    \country{China}
}

\author{Guangming Lu}
\email{glu@linkedin.com}
\affiliation{\institution{LinkedIn Corporation}
    \city{Beijing}
    \country{China}
}

\author{Zi Li}
\email{zili@linkedin.com}
\affiliation{\institution{LinkedIn Corporation}
    \city{Beijing}
    \country{China}
}

\renewcommand{\shortauthors}{Hao Chen et al.}

\begin{abstract}
Online recruitment platforms typically employ Person-Job Fit models in the core service that automatically match suitable job seekers with appropriate job positions. While existing works leverage historical or contextual information, they often disregard a crucial aspect: job seekers' social relationships in professional networks. This paper emphasizes the importance of incorporating professional networks into the Person-Job Fit model. Our innovative approach consists of two stages: (1) defining a Workplace Heterogeneous Information Network (WHIN) to capture heterogeneous knowledge, including professional connections and pre-training representations of various entities using a heterogeneous graph neural network; (2) designing a Contextual Social Attention Graph Neural Network (CSAGNN) that supplements users' missing information with professional connections' contextual information. We introduce a job-specific attention mechanism in CSAGNN to handle noisy professional networks, leveraging pre-trained entity representations from WHIN. We demonstrate the effectiveness of our approach through experimental evaluations conducted across three real-world recruitment datasets from LinkedIn, showing superior performance compared to baseline models.
\end{abstract}


\maketitle

\section{Introduction}

With the rapid development of the Internet, online recruitment platforms (e.g., LinkedIn\footnote{https://www.linkedin.com}{}, Indeed\footnote{https://www.indeed.com}{}, and ZipRecruiter\footnote{https://www.ziprecruiter.com}{}) are becoming essential for recruiting and job seeking. A considerable number of talent profiles and job descriptions are posted on these platforms. Taking LinkedIn as an example, more than 900 million members have registered, and 90 jobs were posted every second by the first quarter of 2023\footnote{https://news.linkedin.com/about-us\#Statistics}{}. Considering such a large number of options, Person-Job Fit (PJF) \cite{malinowski2006matching} has become a critical research topic for improving the efficiency of recruitment and job seeking. Person-Job Fit aims to automatically link the right talents to the right job positions according to talent competencies and job requirements. 

Previous works on Person-Job Fit mainly focus on leveraging two types of information, namely, (1) historical job application information and (2) textual information in profiles and job descriptions. Collaborative filtering-based methods \cite{shalaby2017help,bian2020learning} are applied to capture co-apply relations between job seekers and co-applied relations between job positions in the historical application information. Manually-engineered textual features and deep language models have been widely adopted to leverage the textual information \cite{bian2019domain,qin2018enhancing, yan2019interview, lu2013recommender}.

However, workplace social connections among members, commonly called professional networks, have been overlooked as a pivotal source of information. A survey reported on the Official LinkedIn Blog\footnote{\url{https://blog.linkedin.com/2017/june/22/the-best-way-to-land-your-next-job-opportunity}} showed that out of more than 15,000 LinkedIn members, 80\% believed that professional networking could help find new job opportunities, an even more striking 70\% gained job opportunities directly through their connections. As an important source for achieving employment relationships, professional networks have great potential for helping with Person-Job Fit. Incorporating professional networks into the models can offer two main benefits: (1) Professional networks directly improve Person-Job Fit by bridging the gap between job seekers and potential job opportunities. A Person-Job Fit model enriched with professional network information can suggest job opportunities to users based on their professional connections. (2) Professional networks can also help complete job seekers' profiles. Often job seekers' online profiles lack comprehensive details, yet one's professional experience and skills can be, to some extent, discernible through their professional connections.

While professional networks offer advantages, they often contain a lot of noise. This noise includes connections that aren't relevant and information from relevant connections that don't help improve Person-Job Fit. To illustrate, irrelevant connections might involve job seekers' former classmates or recruiters from unrelated industries. Even when the connections are relevant to the job seeker, they could still have information unrelated to the job seeker's aspirations for the role. To tackle these noises, a promising strategy involves using heterogeneous knowledge, such as job seekers' skills, work experience, and educational background, to gauge professional connections' relevance and extract pertinent information to elevate Person-Job Fit.

In this paper, we propose a graph neural network-based framework that utilizes heterogeneous knowledge to integrate professional networks into Person-Job Fit. We address the challenge of social noise in professional networks by designing a job-specific attention mechanism. Initially, we define a Workplace Heterogeneous Information Network (WHIN) that captures heterogeneous knowledge, including professional connections. We employ a heterogeneous graph pre-training technique to learn the representations of various entities in the WHIN. Subsequently, we introduce the Contextual Social Attention Graph Neural Network (CSAGNN), designed to supplement users' lacking information with contextual information from their professional connections. To tackle the social noise in the workplace social network, we infuse a job-specific attention mechanism into CSAGNN, capitalizing on the pre-trained entity representations from WHIN.

The main contributions of this study are as follows:
\begin{itemize}
    \item To our knowledge, we are the first to define a heterogeneous information network that incorporates heterogeneous knowledge in the Person-Job Fit scenario.
    
    \item We systematically utilize professional networks in a two-stage approach, WHIN pre-training and CSAGNN, to address the Person-Job Fit task.
    
    \item We present a novel Contextual Social Attention Graph Neural Network (CSAGNN) specifically designed to handle noisy professional networks, effectively mitigating the impact of irrelevant information while focusing on related professional connections and contexts for Person-Job Fit.
    
    \item We evaluate our approach on three real-world datasets across diverse industries. Experimental results show that our model outperforms baseline models.
\end{itemize}

\section{related works}

\begin{figure}[t]
    \centering
    \includegraphics[width=0.6\linewidth]{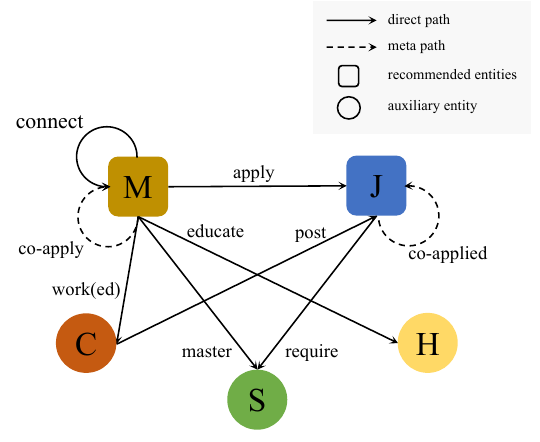}
    \caption{The metagraph of Workplace Heterogeneous Information Network. It encompasses not only members (M) and jobs (J), which are crucial for Person-Job Fit, but also entities such as skills (S), companies (C), and schools (H).}
    \label{fig:metagraph}
\end{figure}

The related work of our study can be grouped into two main categories, namely,  \textit{Person-Job Fit} and \textit{Heterogeneous Information Network-based Recommendation}.

\subsection{Person-Job Fit}

As a core function of the online recruitment platform, Person-Job Fit \cite{malinowski2006matching} has received widespread attention.

Mainstream works view Person-Job Fit as a text-matching problem between member profiles and job descriptions to fully use the rich contextual knowledge. For example, JLMIA \cite{shen2018joint} is a latent variable model to model job descriptions and member profiles jointly. PJFNN \cite{zhu2018person} encodes member profiles and job descriptions by hierarchical CNN. BPJFNN \cite{qin2018enhancing} leverage BiLSTM to get the semantic representation of each word in member profiles and job descriptions. APJFNN \cite{qin2018enhancing} automatically weighs abilities mentioned in textual information based on historical recruitment results. A transferable deep global match network \cite{bian2019domain} is proposed to solve the domain adaptation problem in three levels for Person-Job Fit. JRMPM \cite{yan2019interview} proposes a matching network with preference modeled to explore the latent preference given the history of the matching process.

\begin{figure*}[!ht]
    \centering
    \includegraphics[width=0.75\textwidth]{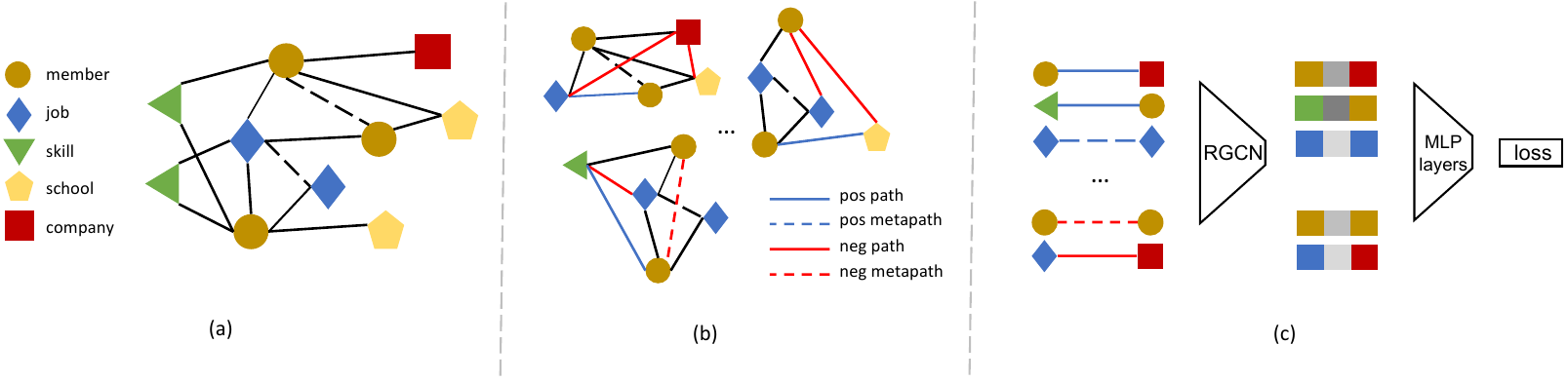}
    \caption{Steps of WHIN pre-training. (a) Workplace heterogeneous graph with metapath. (b) Subgraph sampling for mini-batch pre-training. (c) A pre-training model with encoder-decoder architecture using Link-level pre-training task.}
    \label{fig:WHIN_pretrain}
\end{figure*}

Some works take structure knowledge into consideration. For example, MV-CoN \cite{bian2020learning} adopts a co-teaching mechanism to capture semantic and structure knowledge at the same time. DPGNN \cite{yang2022modeling} explicitly models the two-way selection preference for PJF using GCN. KGRM \cite{yao2022knowledge} model members and jobs as two graphs and fuse prior external knowledge, e.g., skill knowledge graph, into the graph representations.

Previous studies have made good use of textual information, skill entity, and direct interaction between members and jobs to help with Person-Job Fit. However, these studies often overlook the importance of workplace social relations and the incorporation of diverse, heterogeneous knowledge.

\subsection{Heterogeneous Information Network-based Recommendation}

Recommender systems have been widely deployed on the Internet to alleviate information overload \cite{jacoby1984perspectives}. Due to the excellent ability to model complex auxiliary information, Heterogeneous Information Network (i.e., HIN) has become one of the mainstream approaches in recommendation tasks. Depending on the training method, HIN-based models can be divided into \emph{two-stage training-based models} and \emph{end-to-end training-based models} \cite{liu2022survey}.

Two-stage training-based models learn low-dimensional representations of nodes and graph structure using unsupervised tasks and use these representations on various downstream tasks \cite{du2022understanding,bi2022make,huang2023robust,bi2023mm}. Inspired by DeepWalk \cite{perozzi2014deepwalk} and node2Vec \cite{grover2016node2vec}, metapath2vec \cite{dong2017metapath2vec} leverages metapath-based random walks to construct a heterogeneous neighborhood of nodes and leverages skip-gram model to generate node representations. Similarly, HIN2Vec \cite{fu2017hin2vec} defines an unsupervised metapath prediction task and jointly learns metapath predictors and node representations. Many models are proposed based on the graph representation methods such as those above. For example, HERec generates node sequences with metapath-based random walks, uses node2vec to learn node representations, and completes recommendations by representation similarity. MAGNN \cite{fu2020magnn} leverages node content features and information of intermediate nodes along the metapath by node content transformation, intra- and inter-metapath aggregation. Besides, there are some methods not relying on manual metapath, such as HRLHG \cite{jiang2018cross}, NREP \cite{yang2019citation}, and ECHCDR \cite{li2020heterogeneous}.

Compared with two-stage training-based models, end-to-end-based models can use supervision signals directly while training. Thus, they are more customized to specific downstream tasks. For HINs with rich relations, relation-aware graph neural networks can achieve great results. For example, RGCN \cite{schlichtkrull2018modeling} assigns a weight matrix to each type of relation, thus extending GCN \cite{welling2016semi} to multi-relation graphs. DisenHAN \cite{wang2020disenhan} projects nodes into different subspaces using type-specific transformation and uses intra- and inter-aggregation to learn node representations. HAN \cite{wang2019heterogeneous} uses a dual attention mechanism to aggregate information from different metapaths. HPN \cite{ji2021heterogeneous} designs a semantic fusion mechanism for learning the importance of metapath and fusing them judiciously. By using only its immediate connections as input, HGT \cite{hu2020heterogeneous} can learn and extract relevant metapaths for various tasks through its automatic and implicit attention mechanism without requiring manual metapath design.

In this paper, we adopt the two-stage training-based paradigm. During the pre-training stage, we integrate heterogeneous knowledge specific to the Person-Job Fit scenario to obtain diverse entity representations that encompass both structural and textual information. In the downstream stage, we design a novel model that leverages the information from the pre-training stage to filter out noise within professional networks. As a point of comparison, we have also selected two end-to-end-based models as baselines.

\section{Methodlogy}

\begin{figure*}[!ht]
    \centering
    \includegraphics[width=0.75\textwidth]{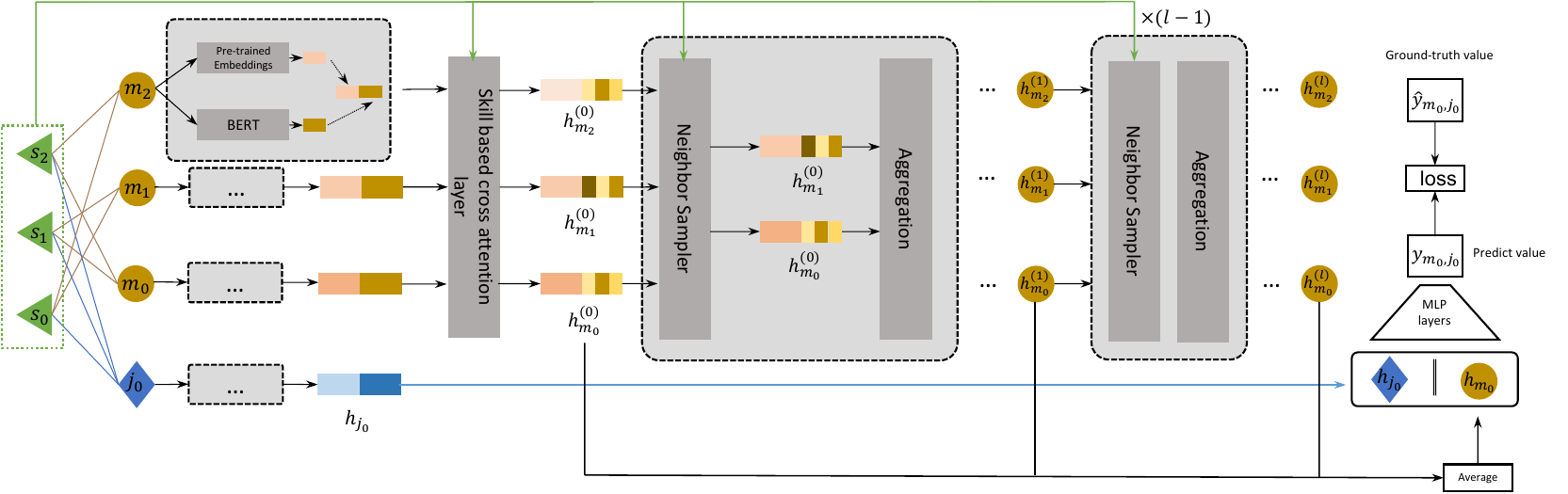}
    \caption{Architecture of CSAGNN. When determining the match between job $j_0$ and member $m_0$, the information from $m_0$'s professional connections, $m_1$ and $m_2$, is simultaneously acquired. The initial representations of both the member and job are formed by concatenating the WHIN pre-training embedding with the representation obtained after processing the text information through BERT. All member text information is re-weighted through an attention mechanism based on the skills $s_i$ required by $j_0$. A neighbor sampler module, operating based on the similarity between different members and the required skills, can filter out professional connections that are not relevant to the job.}
    \label{fig:CSAGNN_framework}
\end{figure*}

Here we introduce our two-stage approach to improve the performance of the Person-Job Fit task by incorporating professional networks. The task and notation are formally defined in Section \ref{Formal Task Description and Notations}. In the first stage, we pre-train on the Workplace Heterogeneous Information Network (WHIN), which is detailed in Section \ref{The Construction of Workplace Heterogeneous Information Network}. We introduce the pre-training approach on WHIN in Section \ref{Pre-training on Workplace Heterogeneous Information Network}. To overcome the noise in professional networks, we propose the contextual social attention graph neural network (CSAGNN) in the second stage, as described in Section \ref{Contextual Social Attention Graph Neural Network}.

\subsection{Formal Task Description and Notations}
\label{Formal Task Description and Notations}
Person-Job Fit aims to match a set of job seekers with a set of job opportunities. Assume that there is a set of members $\mathcal{M}=\set*{m_1, m_2,\dots , m_n}$ who may be seeking job opportunities and a set of jobs $\mathcal{J}=\set*{j_1, j_2, \dots, j_m}$. Formally, given a candidate pair $(m_j, j_k)$, the model is required to learn a function $\mathcal{F}$ to predict the probability of whether the member applied for the job.
\[
    \mathcal{F}(m_i, j_k) =  \begin{cases}
        1 & m_i \text{ applied for } j_k \\
        0 & m_i \text{ didn't apply for } j_k
    \end{cases} \\
\]
The set of candidate pairs can be defined as

$$T=\set*{(m_i, j_k)|m_i\in \mathcal{M}, j_k\in \mathcal{J}}$$.

In addition to the sets of members ($\mathcal{M}$) and jobs ($\mathcal{J}$), there exist three other sets of entities that are strongly related to members and jobs: skills ($\mathcal{S}$), companies ($\mathcal{C}$), and schools ($\mathcal{H}$). These entities are referred as auxiliary information and can be leveraged to improve the model's accuracy. Given an entity $e$ and a relation $r$, an interaction map $\mathcal{A}_{e}^{r}$ is used to specify the destination entities related to $e$ through $r$. For example, $\mathcal{A}_{m_i}^{apply}$ represents the set of jobs applied by $m_i$. If $m_i$ has not applied for any jobs, $\mathcal{A}_{m_i}^{apply}=\emptyset$.

\subsection{The Construction of Workplace Heterogeneous Information Network}
\label{The Construction of Workplace Heterogeneous Information Network}

The Person-Job Fit scenario contains a wide range of heterogeneous knowledge, including various entities such as skills, companies, and schools, as well as relationships such as professional connections and skill requirements. We define a Workplace Heterogeneous Information Network (WHIN) to model this knowledge. As shown in Figure \ref{fig:metagraph}, WHIN is a heterogeneous information network consisting of five types of entities, namely Member, Job, Skill, Company, and School. It also contains nine types of relations, which are introduced to capture the different types of connections between the entities.
 
The Workplace Heterogeneous Information Network primarily consists of natural relations among diverse entities. Besides, two additional metapaths have been artificially constructed. The first metapath, co-apply ($\mathcal{M}$-$\mathcal{J}$-$\mathcal{M}$), connects members who have applied for the same job. The second metapath, co-applied ($\mathcal{J}$-$\mathcal{M}$-$\mathcal{J}$), connects multiple jobs that have been applied for by the same member. These metapaths were constructed based on prior knowledge and have proven to be useful in enhancing the performance of the WHIN pre-training process \cite{fu2020magnn}.

\subsection{Pre-training on Workplace Heterogeneous Information Network}
\label{Pre-training on Workplace Heterogeneous Information Network}

\subsubsection{link-level pre-training task}
\label{link-level pre-training task}
We utilize a heterogeneous graph neural network with a link-level pre-training task to acquire node representations with rich structural knowledge. To improve the scalability of the pre-training process, we create mini-batches by sampling subgraphs from the entire Workplace Heterogeneous Information Network (WHIN), as illustrated in Figure \ref{fig:WHIN_pretrain} (a) and (b).

Specifically, a batch of candidate pairs is selected as the source nodes. From these source nodes, their k-hop neighbors are sampled to construct a subgraph based on each direct relation or metapath. These identified relations and metapaths serve as positive samples. In addition, to introduce negative samples, we randomly select pairs of entities that do not exhibit the expected relationships. The link-level pre-training task revolves around determining the existence of the given samples. We employ an encoder-decoder structure, as illustrated in Figure \ref{fig:WHIN_pretrain} (c). The encoder learns node representations within each subgraph, while the decoder predicts link existence.

\subsubsection{pre-training model structure}
\label{pre-training model structure}

During the pre-training phase, we select a fixed number of tokens from the textual content of each entity type. These selected tokens are then processed through the BERT model \cite{devlin2018bert}. The resulting BERT outputs are averaged to create initial representations for each entity.

In a formal context, when working with a textual representation associated with an entity denoted as $e_i\in \mathcal{E}$, we construct a corresponding initial embedding referred to as $z_i^{(0)}$ by utilizing BERT. Specifically, We leverage member profiles, job descriptions, skill names, and other textual descriptions to initialize entity representations, thereby enhancing the initial embeddings for each entity type.

After obtaining an initialized representation with semantic information, we utilize an encoder, RGCN \cite{schlichtkrull2018modeling}, to integrate the information between entities. The message-passing process in the RGCN encoder is represented by the following equation:

\begin{equation}
    z_i^{(l+1)}=\sigma\left(\sum_{r\in R}\sum_{j\in {\mathcal{A}_{e_i}^{r}}}\frac{1}{c_{i,r}} W_r^{(l)} z_j^{(l)} + W_0^{(l)}z_i^{(l)}\right),
\end{equation}

where $W_r^{(l)}$ is a parameter matrix at the $l$-th layer of relation $r$, $W_0^{(l)}$ is used for self-loop at the $l$-th layer and $c_{i, r} = \left|\mathcal{A}_{e_i}^{r}\right|$ is a normalization constant.

We employ $L$ encoder layers for Link Prediction, utilizing the final representation $z_i^{(L)}$ for entity $e_i$. We apply a multi-layer perceptron (MLP) to score each relation or metapath $r_i$, represented by vector $M_{r_i}$. The score for a particular link $(s, r, d)$ is determined as follows:

\begin{equation}
y_{(s, r, d)}= MLP(z_s||M_r||z_d)
\end{equation}

We employ cross entropy as the loss function for optimizing the pre-training model. Notably, the described pre-training process incorporates professional network information into entity embeddings.

\begin{equation}
    \label{eqn: pretrain_loss}
    \mathcal{L}_{pretrain}=-\left(y_{(s, r, d)}\log y'_{(s, r, d)}+\left(1-y_{(s, r, d)}\right)\log\left(1-y'_{(s, r, d)}\right)\right)
\end{equation}

\subsection{Contextual Social Attention Graph Neural Network}
\label{Contextual Social Attention Graph Neural Network}

As discussed earlier, professional networks are inherently noisy. To capture job-specific insights from the profiles of $m_i$'s professional connections, a multi-head attention (MHA) mechanism \cite{vaswani2017attention} is employed. As shown in Eq. \ref{eqn: jk-specific contextual representation}, $c_i$ is the profile representation of member $m_i$ obtained from a pre-trained BERT model. The MHA mechanism uses each required skill embedding $z_{s_k}$ as the query, and $c_i$ as both the key and value. It computes attention weights between $c_i$ and the required skill embeddings. The weighted sum averages these attention outputs for all required skills. This results in the job-specific contextual feature $F_{m_i}^c(j_k)$ that captures relevant information from $m_i$'s connections for job $j_k$.

\begin{equation}
\label{eqn: jk-specific contextual representation}
F_{m_i}^c(j_k) = \frac{1}{\left |\mathcal{A}_{j_k}^{require}\right |}\sum_{s_k \in \mathcal{A}_{j_k}^{require}}{
MHA(Q=z_{s_k}, K=c_i, V=c_i)
}
\end{equation}

As shown in Eq. \ref{eqn:initial feature h}, contextual feature $F_{m_i}^c(j_k)$ is concatenated with the WHIN pre-trained feature $F_{m_i}^s$ to form member's $j_k$-specific initial feature.

\begin{equation}
    \label{eqn:initial feature h}
    h_{m_i}^{(0)}(j_k) = (F_{m_i}^c(j_k) || F_{m_i}^s)
\end{equation}

The relevance degree of professional connections to the job $j_k$ needs to be considered in the messaging process. Here, we use the distance between WHIN pre-trained representations of skills mastered by members and the skills required by jobs as the weights during message passing.

\begin{equation}
    \label{eqn: average skill embedding}
    z'({\hat{S}}) = {\sum_{s \in \hat{S}}{z_s}}/|\hat{S}|
\end{equation}

\begin{equation}
    \label{eqn: distance between skills}
    d_{m_i}(j_k) = \frac{z'(\mathcal{A}_{m_i}^{master})\cdot z'(\mathcal{A}_{j_k}^{require})}{\sqrt{(z'(\mathcal{A}_{m_i}^{master}))^2}\sqrt{(z'(\mathcal{A}_{j_k}^{require}))^2}}
\end{equation}

% d_{m_i}(j_k) = \dvert*{z'(\mathcal{A}_{m_i}^{master})-z'(\mathcal{A}_{j_k}^{require})}\\
Given a set of skills $\hat{S}$, average embedding can be described in Eq. \ref{eqn: average skill embedding} and distance between members' and jobs' skills $d_{m_i}(j_k)$ can be described in Eq. \ref{eqn: distance between skills}. Further, due to some members or jobs may have hundreds of related skills, we sampled the skill set like $\mathcal{A}_{m_i}^{master}$ and $\mathcal{A}_{j_k}^{require}$ in Eq. \ref{eqn: jk-specific contextual representation} and Eq. \ref{eqn: distance between skills}. The number of sampled skills, $n_s$, is a hyperparameter of CSAGNN. We will discuss the influence of $n_s$ in Section \ref{RQ2: Ablation Study}.

\begin{equation}
    \label{eqn: social message passing}
    h_{m_i}^{(l+1)}(j_k) = \sigma {\left(W_{2}^{(l)}\sum_{m'\in \mathcal{A}_{m_i}^{connect}}\alpha_{m_j}(j_k)h_{m_j}^{(l)}(j_k)+W_1^{(l)} h_{m_i}^{(l)}\right)}
\end{equation}

\begin{equation}
    \label{eqn: softmax weight}
    \alpha_{m_j}(j_k) = \frac{d_{m_j}(j_k)}{\sum_{m'\in \mathcal{A}_{m_i}^{connect}}{d_{m'}(j_k)}}
\end{equation}

As shown in Eq. \ref{eqn: social message passing}, we can get the messaging function of professional network enhanced GNN by text attention and skill distance attention where $\alpha_{m_j}(j_k)$ is normalized message aggregation weight described in Eq. \ref{eqn: softmax weight}. We use the average of each layer's representation as the final representation of $m_i$:

\begin{equation}
\label{eqn: residual sum}
    h_{m_i}(j_k) = \sum_{l=0}^{L+1}{h_i^{(l)}{j_k}}/L
\end{equation}

Compared with member profiles, job descriptions provided by recruiters often have a high quality. Thus, we directly use contextual features and WHIN pre-trained features as the final representation for all job entities.

\begin{equation}
\label{eqn: job init representation}
    h_{j_k} = (F_{j_k}^c || F_{j_k}^s)
\end{equation}

$h_{j_k}(m_i)$ and $h_{m_i}(j_k)$ will be concatenated and passed through a vanilla multi-layer Perceptron (MLP) to get the predicted value $y_{m_i, j_k}$, and the cross entropy loss could be formulated as Eq. \ref{eqn: loss of CSAGNN}, where $y_{m_i,j_k}'$ is the ground truth:

\begin{equation}
    y_{m_i, j_k} = MLP(h_{m_i}(j_k)||h_{j_k})
\end{equation}

\begin{equation}
    \label{eqn: loss of CSAGNN}
    \mathcal{L}_{csa}= -\left(y_{m_i, j_k}\log y'_{m_i, j_k} + \left(1-y_{m_i, j_k}\right)\log\left(1-y'_{m_i, j_k}\right)\right)
\end{equation}

\begin{table*}[ht]
    \caption{Performance comparison of all baselines and our models. The best and second-best results are shown in bold and underlined, respectively. All values are multiplied by 100.}
    \centering
    \includegraphics[width=.95\linewidth]{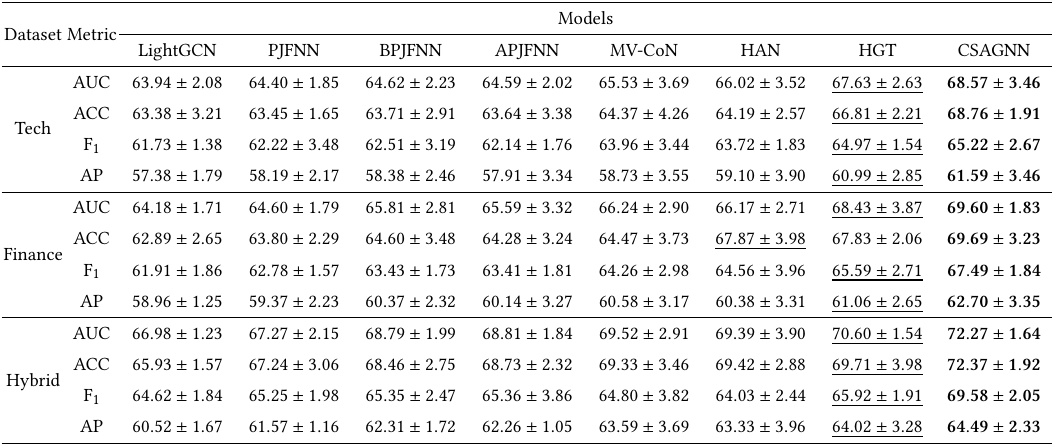}
    \label{tab:Performance}
\end{table*}

\section{Experiments}

In this section, we will validate the performance of the CSAGNN model on three real datasets and answer the following research questions:

\begin{itemize}
    \item \textbf{RQ1}: Can CSAGNN outperform existing models on real-world datasets?
    \item \textbf{RQ2}: What are the effects of different components in our model?
    \item \textbf{RQ3}: Are professional networks helpful for the Person-Job Fit task?
    \item \textbf{RQ4}: Does the WHIN pre-training provide meaningful representations of skills for CSAGNN?
\end{itemize}

\subsection{Experiment Setup}

\subsubsection{Datasets}

We validate our model on three large real-world datasets from LinkedIn, a well-known workplace social platform. Datasets are divided by the industry to which the members and jobs belong. The first two datasets contain members and jobs from the technology and finance industries. Due to the singularity of the industry, social noise is limited. The third dataset, which includes members and jobs from various industries such as healthcare, education, and semiconductors, is intended to test the ability of CSAGNN to filter noise in professional networks. This hybrid dataset implies more complex social relationships and a wider range of job opportunities. The WHINs constructed from these three datasets all contain about 200,000 entities and 10 million links. Table \ref{tab:dataset statistic} displays the statistics of the datasets. The skills, schools, and companies in the dataset all come from explicit annotations by members or jobs on LinkedIn. We cannot provide further detailed information about the datasets to protect the users' privacy. Each candidate pair within the datasets is labeled as either positive or negative based on the feedback provided by members.

\begin{table}[ht]
  \caption{Statistics of datasets. M refers to Member, J refers to Job, S refers to Skill, CP refers to Candidate Pair, and PC refers to Professional Connection.}
  \label{tab:dataset statistic}
  \begin{tabular}{cccccc}
    \toprule
    \textbf{Industry} & \textbf{\#M} & \textbf{\#J} & \textbf{\#S} & \textbf{\#CP} &\textbf{\#PC}\\
    \midrule
    Tech        &       33000+   &       62000+ &  27000+ &   136000+  &   1922000+       \\
    Finance     &       20000+   &       27000+    & 23000+ & 36000+  &   615000+       \\
    Hybrid         &       83000+   &       120000+  & 33000+  &   200000+  &   2768000+       \\
    \bottomrule
  \end{tabular}
  % }
\end{table}

\subsubsection{Baseline Models}

We compare our model with the following baseline models.

\begin{itemize}
    \item \textbf{LightGCN \cite{he2020lightgcn}} is a simplified Graph Convolutional Network model for collaborative filtering with competitive performance and less complexity.
    \item \textbf{PJFNN \cite{zhu2018person}} is a method based on a convolutional neural network (CNN). Hierarchical CNN encodes Resumes and job descriptions independently, and the matching degree is calculated by cosine similarity.
    \item \textbf{BPJFNN \cite{qin2018enhancing}} leverages bidirectional LSTM to learn the representations of resumes and job descriptions. 
    \item \textbf{APJFNN \cite{qin2018enhancing}} leverages bidirectional LSTM and hierarchical attention mechanism to learn the representations of resumes and job descriptions.
    \item \textbf{MV-CoN \cite{bian2020learning}} combines text matching model and RGCN to learn representations of resumes and job descriptions.
    \item \textbf{HAN \cite{wang2019heterogeneous}} uses a dual attention mechanism to aggregate neighbor information via different metapaths.
    \item \textbf{HGT \cite{hu2020heterogeneous}} designs node- and edge-type dependent parameters to characterize the heterogeneous attention over each edge.
\end{itemize}

These baselines can be divided into three groups: (1) context-based models that treat Person-Job Fit as a text match problem and use contextual knowledge from members' profiles and jobs' descriptions: PJFNN, BPJFNN, APJFNN, and MV-CoN. In particular, MV-CoN additionally introduces structural information to enhance model performance. (2) collaborative filtering-based model that uses direct interactions between members and jobs: LightGCN. (3) end-to-end heterogeneous graph neural network models: HAN, HGT.

\subsubsection{Evaluation and Implementation Details}
We use four widely used metrics to evaluate the ranking performance: AUC, accuracy (ACC), F1, and average precision (AP).

The Person-Job Fit models, namely PJFNN, BPJFNN, and APJFNN, are implemented using RecBole—an established open-source recommendation library \cite{zhao2021recbole}. The MV-CoN model leverages the original code provided in their respective paper \cite{bian2020learning}. Other models are implemented with PyTorch Geometric \cite{Fey/Lenssen/2019}. We have employed BERT-Tiny\footnote{\url{https://huggingface.co/prajjwal1/bert-tiny}} to reduce computational costs.

The embedding dimensions for all models are standardized at 32. As discussed in Section \ref{link-level pre-training task}, subgraph sampling for mini-batch WHIN pre-training is executed using the PyTorch Geometric subgraph sampler \cite{Fey/Lenssen/2019}. Subgraph construction starts from nodes within a batch of candidate pairs, and each hop samples five neighbors via relations (or metapaths). Reconnecting all relations (and metapaths) between the sampled nodes follows this. We set the number of sampled hops to 3. The number of sampled hops is fixed at 3. In CSAGNN, the count of sampled skills ($n_s$) is designated as 10, and CSAGNN comprises two layers. The first 128 tokens from all text inputs are captured for all models. All models are optimized using the Adam optimizer \cite{kingma2014adam}, with a learning rate adjusted from 0.01 to 0.0001. For the purpose of evaluation, the dataset utilized across all models is randomly partitioned into a ratio of 8:1:1 for training, validation, and testing respectively. Three independent experiments are conducted, each one repeated, to ascertain consistent and reliable outcomes.

\subsection{The Overall Comparison (RQ1)}

Table \ref{tab:Performance} shows the performance of all baseline models and our model, CSAGNN. The results indicate better performance on the hybrid dataset than on individual industry datasets, possibly because jobs within the same industry are more similar and thus harder to distinguish. End-to-end heterogeneous graph neural network models have shown superior performance on all three datasets compared to context-based and CF-based models. HGT, utilizing heterogeneous knowledge, has improved AUC scores by 3.20\%, 3.31\%, and 1.55\% compared to the best baseline MV-CoN, which employs homogeneous knowledge. CSAGNN further improved AUC scores by 1.39\%, 1.71\%, and 2.37\% compared to HGT.

\begin{table}[ht]
    \caption{Ablation studies conducted on all datasets, with all values multiplied by 100.}
    \label{tab:ablation}
    \centering
    \includegraphics[width=0.95\linewidth]{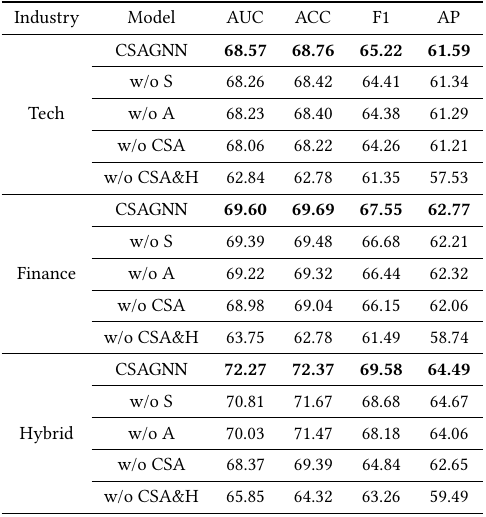}
\end{table}

\begin{figure}[ht]
    \centering
    \subfigure[Model performance varying the number of sampled skills while fixing CSAGNN layers to 1.]{
        \includegraphics[width=.32\linewidth]{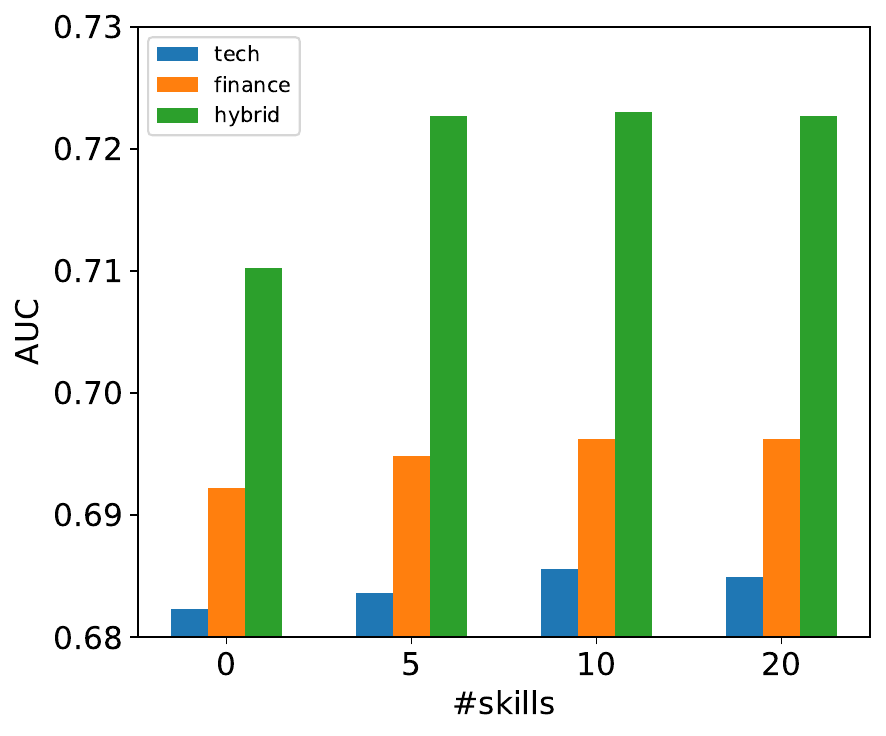}
        \hspace{0.2cm}
        \includegraphics[width=.32\linewidth]{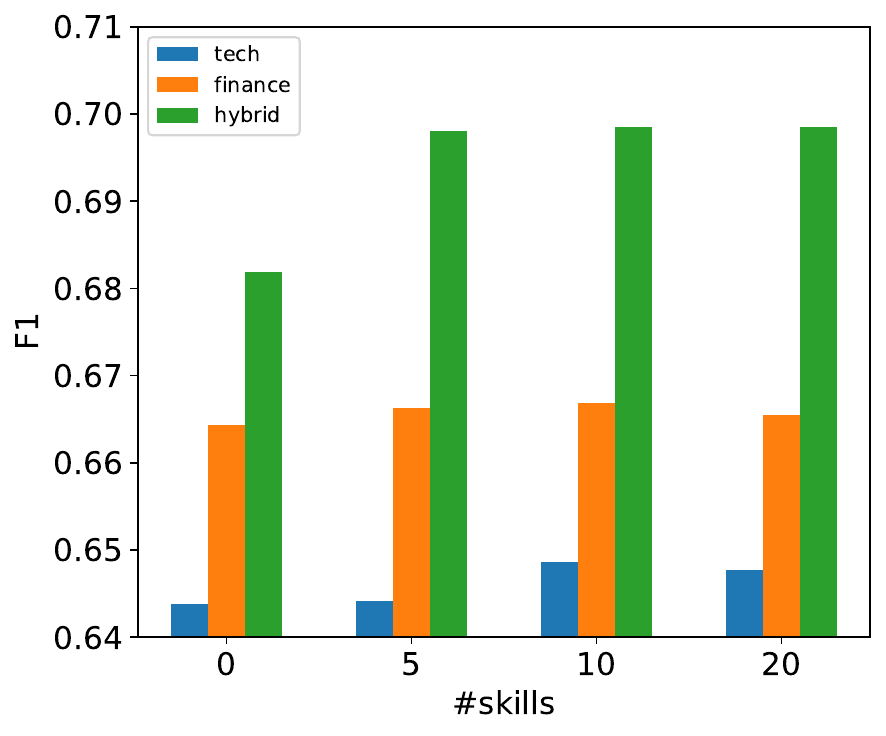}
    }
    \\
    \subfigure[Model performance varying the number of CSAGNN layers while fixing the number of sampled skills to 10.]{
        \includegraphics[width=.32\linewidth]{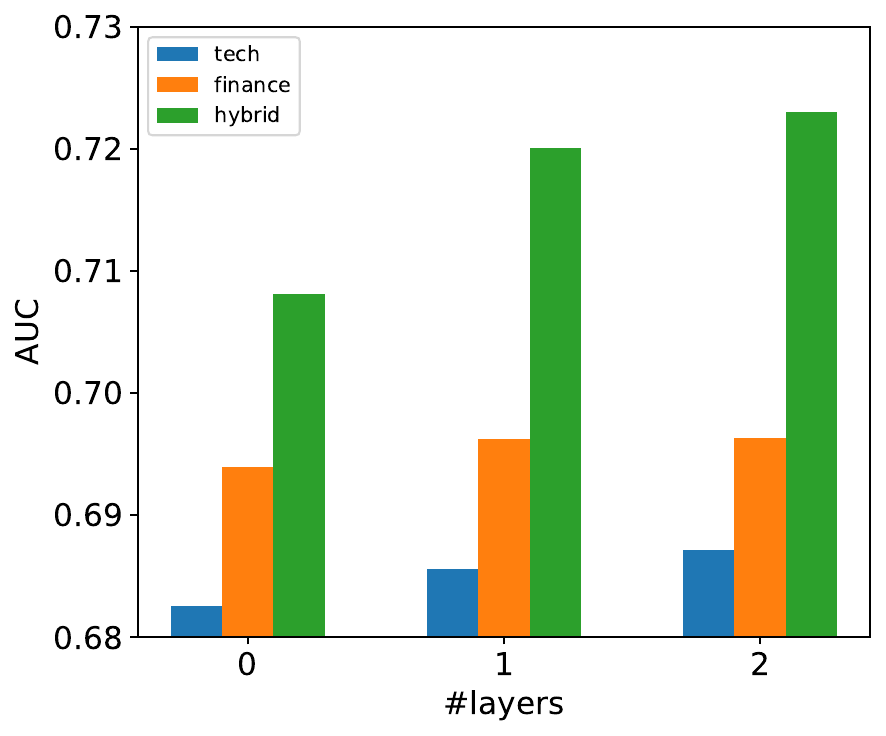}
        \hspace{0.2cm}
        \includegraphics[width=.32\linewidth]{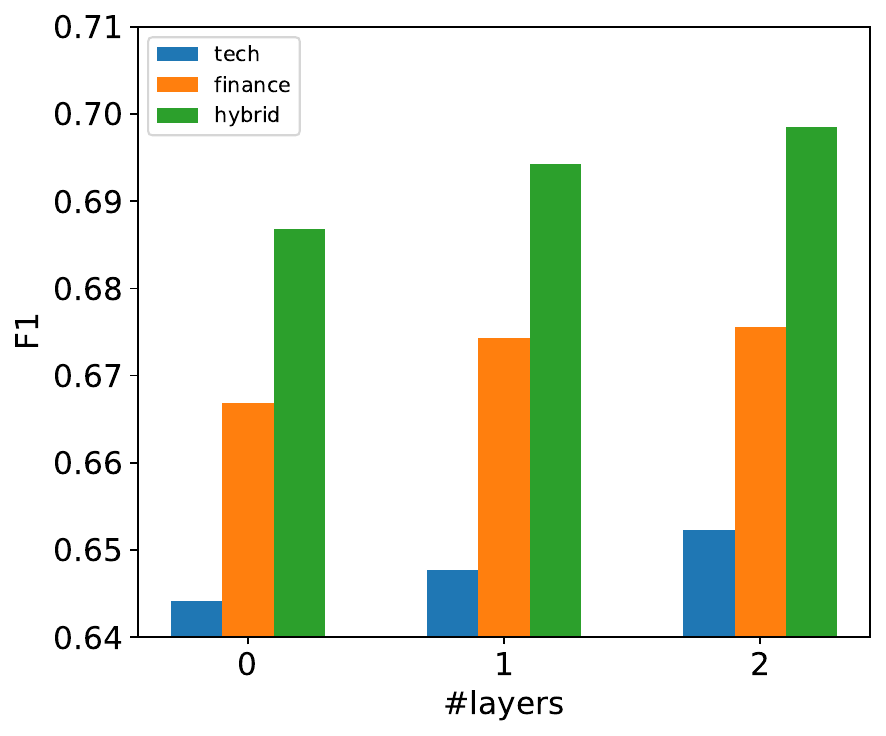}
    }
    \caption{Hyperparameter tuning experiments to investigate the specific effects of social relations and job-specific attention mechanism.}
    \label{fig:skill_range}    
\end{figure}

\subsection{Ablation Studies and  Parameter Tuning (RQ2)}
\label{RQ2: Ablation Study}
Our approach employs two main techniques: WHIN pre-training, which integrates heterogeneous knowledge, including professional networks, and CSAGNN, which incorporates professional networks with an attention mechanism. We conducted ablation studies to analyze the effectiveness of different techniques, where we considered the following variants of CSAGNN: (1) CSAGNN w/o S removes messages from professional connections but retains the job-specific attention mechanism for members themselves. (2) CSAGNN w/o A removes the job-specific attention mechanism. (3) CSAGNN w/o CSA removes both professional connections and job-specific attention mechanisms, using only structural knowledge from the WHIN pre-train approach. (4) CSAGNN w/o CSA\&H removes WHIN pre-trained embeddings and CSA mechanism, only using text information as input and leverages MLP to predict.

\begin{figure*}[ht]
    \centering
    \includegraphics[width=0.53\textwidth]{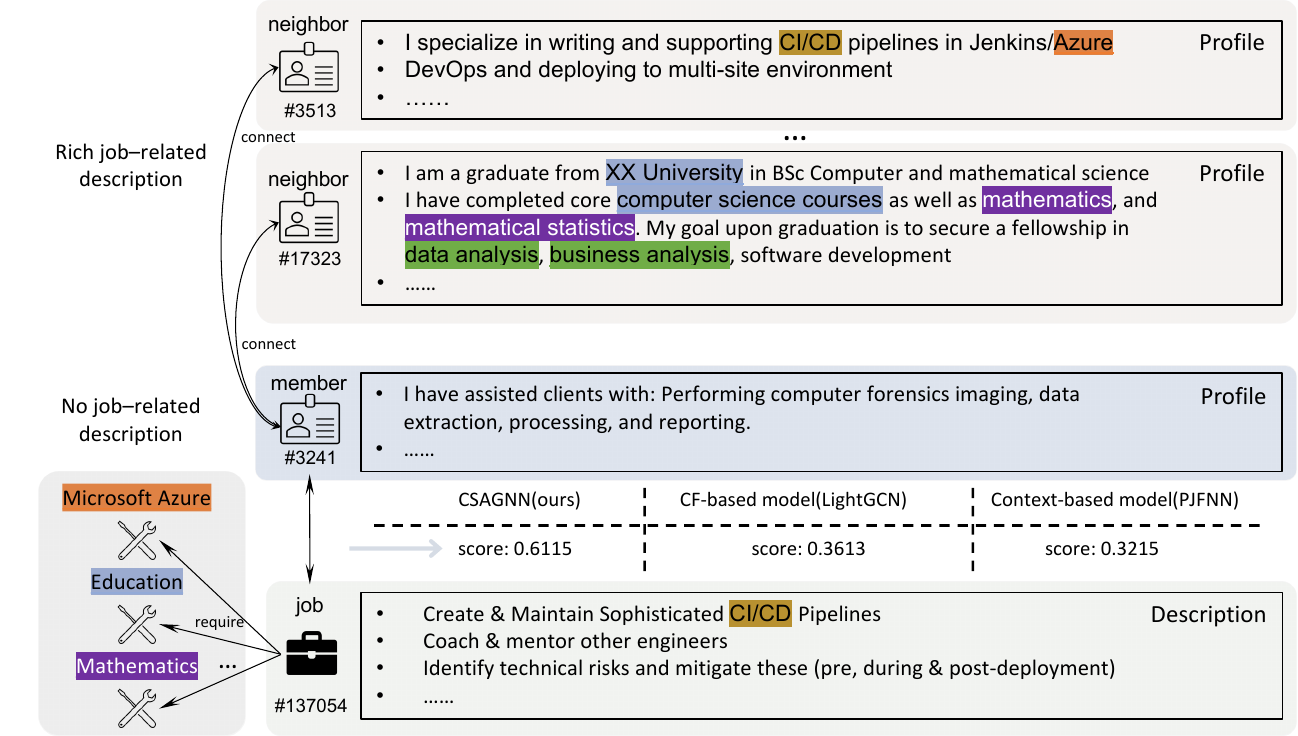}
    \caption{A case where professional connections improve performance on Person-Job Fit. CSAGNN can improve the performance of Person Job Fit by filtering and aggregating information from professional networks. For privacy protection reasons, we have rewritten the statement in the example while ensuring that the semantic information remains unchanged.}
    \label{fig:case}
\end{figure*}

\begin{figure}[ht]
    \centering
    \includegraphics[width=.45\linewidth]{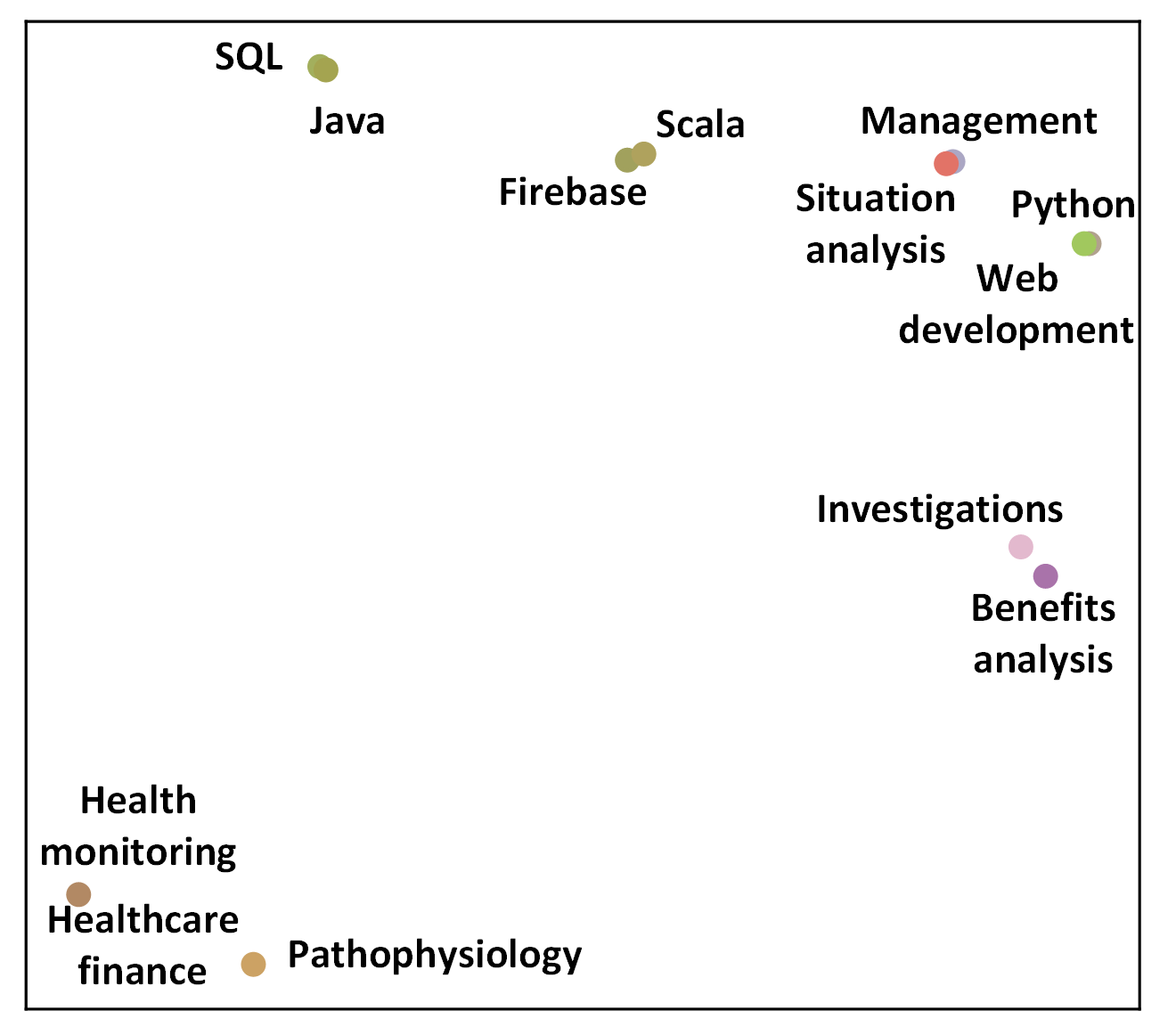}
    \caption{Visualization of WHIN Pre-trained skill embeddings showing clear distinction between Technology and Health-Related Skills.}
    \label{fig:skill_group}
\end{figure}

As detailed in Table \ref{tab:ablation}, within single industry datasets with low social noise, the job-specific attention mechanism and professional connections' messages yielded limited benefits. However, WHIN pre-training embeddings were crucial for enhancing performance. On hybrid datasets with high social noise, where candidate pairs may span different industries, the job-specific attention mechanism, along with the WHIN pre-training embedding, significantly improved the model.

To investigate the specific roles of professional networks and job-specific attention mechanisms in our model, we performed hyperparameter tuning experiments by fixing one module and varying the other to observe its impact on performance. Specifically, we first fixed the number of CSAGNN layers to 1 and tested the model's performance with different numbers of sampled skills ($n_s$), as described in Section \ref{Contextual Social Attention Graph Neural Network}. We then fixed $n_s$ to 10 and tested the model with different numbers of CSAGNN layers. The results on all datasets showed that, compared to the neighbor sampling range, the gain in model performance from the attention mechanism saturates when the number of sampled skills is relatively small. This observation led us to explore the balance between model performance and efficiency by selecting a smaller value for $n_s$.

\subsection{Case Study (RQ3)}
The value of utilizing professional networks is illustrated in the example shown in Figure \ref{fig:case}. In this case, we have rephrased the information to preserve the users' privacy while retaining its essential meaning. Even though the member's profile contains minimal information, the CSAGNN model adeptly predicts their classification by harnessing job-related insights from the member's professional connections. This contrasts sharply with the incorrect predictions made by the collaborative filtering-based LightGCN and context-based PJFNN models. This specific example accentuates the vital role of professional connections in the classification process and underlines the CSAGNN model's distinctive capability to leverage such relationships for precise predictions.

\subsection{Visualization (RQ4)}
To demonstrate the effectiveness of WHIN pre-training in capturing meaningful skill representations for CSAGNN, we analyzed the pre-trained embeddings of selected skills in the embedding space. Utilizing Principal Component Analysis (PCA) as our method of dimensional reduction \cite{jolliffe2016principal}, we were able to visualize the WHIN pre-trained embeddings for various skills. Figure \ref{fig:skill_group} reveals that skills closely related to programming, such as C++ and Python, are grouped together in the embedding space. In contrast, skills associated with healthcare form a distinct cluster. It is noteworthy that these classifications were achieved using only the skill names as initialization. Yet, our WHIN pre-training method successfully distinguishes between skill categories by learning from heterogeneous knowledge, highlighting the capability of WHIN to recognize and differentiate skills across multiple domains.

\section{Conclusion}
This paper introduces a novel two-stage approach for leveraging professional networks in Person-Job Fit, including the formation of the Workplace Heterogeneous Information Network (WHIN). WHIN encompasses various entities such as members, jobs, skills, companies, and schools, with an emphasis on the professional connections among members. By employing heterogeneous graph pre-training techniques, the approach acquires representations that integrate professional connections and other diverse information for different entities. These representations are subsequently applied to the CSAGNN model, helping to filter out social noise.

Experimental results show that professional connections provide valuable job-specific insights. The WHIN pre-training method is also promising for applications like skill completion and professional connection recommendations, opening new research avenues. Furthermore, for large-scale applications of CSAGNN, a key area of future work involves reducing the model's computational overhead \cite{lin2020pagraph, li2020compression, li2021trace}.

\begin{acks}
Thanks to Yihan Cao and Yushu Du for their insightful discussion. Thanks to Sriram Vasudevan and Peide Zhong for their comprehensive review feedback.
\end{acks}

\section*{Ethical Considerations}

The integration of professional networks into Person-Job Fit models offers significant potential for enriching recommendations. However, we must recognize and address two vital ethical considerations that may lead to adverse societal implications:

Firstly, the core of our approach involves accessing members' professional connections and corresponding profiles, which raises privacy concerns. In our research, we ensured that the data collection and processing respected privacy by adhering to proper consent mechanisms and limiting access to pertinent information. Future users should follow suit, being mindful of the need for clear and voluntary consent from members and carefully controlling access to connection information.

Additionally, connections often exhibit demographic clustering, which could introduce biases favoring certain groups within networks. The models might infer sensitive attributes like race, gender, or age. Future users should conduct audits to ensure demographic equity and prohibit direct utilization of protected class information.

In summary, while the enhancement of Person-Job Fit models through professional network data brings advancements, it also introduces risks concerning privacy and fairness. These ethical challenges call for a concerted effort from platforms to resolve, ensuring that the innovations foster an environment that is both respectful of individual rights and free from discriminatory biases.

\appendix
\bibliographystyle{ACM-Reference-Format}
\balance
\bibliography{ref}

\end{document}